\documentclass[12pt,letterpaper]{article}
\pdfoutput=1
\usepackage{jheppub}
\usepackage{verbatim}
\usepackage{slashed}

\usepackage{graphicx}
\usepackage{subfigure}
\usepackage{amsthm}

\newcommand{\labell}[1]{\label{#1}}
\def\({\left(} \def\){\right)}
\def\[{\left[} \def\]{\right]}
\def\al{\alpha} \def\bt{\beta}
\def\del{{\partial}}

\def\M{\mathcal{M}}

\newcommand{\non}{\nonumber \\}

\newcommand{\be}{\begin{equation}}
\newcommand{\ee}{\end{equation}}
\newcommand{\bea}{\begin{eqnarray}}
\newcommand{\eea}{\end{eqnarray}}
\newcommand{\ba}{\begin{eqnarray}}
\newcommand{\ea}{\end{eqnarray}}

\newcommand{\beq}{\begin{equation}}
\newcommand{\eeq}{\end{equation}}
\newcommand{\beqa}{\begin{eqnarray}}
\newcommand{\eeqa}{\end{eqnarray}}
\newcommand{\beqar}{\begin{eqnarray*}}
\newcommand{\eeqar}{\end{eqnarray*}}

\newcommand{\reef}[1]{(\ref{#1})}

\newcommand{\eg}{{\it e.g.,}\ }

\newcommand{\mt}[1]{\textrm{\tiny #1}}

\newcommand{\veps}{\varepsilon}

\newcommand{\A}{\mathcal{A}}

\newcommand{\C}{\mathcal{C}}

\newcommand{\R}{\mathcal{R}}

\newcommand{\ha}{{a}}
\newcommand{\hb}{{ b}}
\newcommand{\hc}{{c}}
\newcommand{\hd}{{d}}
\newcommand{\he}{{e}}

\newcommand{\hO}{\mathcal{O}}

\title{Entanglement Entropy for Relevant and Geometric Perturbations}

\author[a]{Vladimir Rosenhaus}
\affiliation[a]{Kavli Institute for Theoretical Physics, \\
 University of California, Santa Barbara, CA 93106}
\emailAdd{vladr@kitp.ucsb.edu}

\author[b]{and Michael Smolkin}
\affiliation[b]{Center for Theoretical Physics and Department of Physics,\\
 University of California, Berkeley, CA 94720}
\emailAdd{smolkinm@berkeley.edu}

\abstract{We continue the study of entanglement entropy for a QFT through a perturbative expansion of the path integral definition of the reduced density matrix. The universal entanglement entropy for a CFT perturbed by a relevant operator is calculated to second order in the coupling. 
We also explore the geometric dependence of entanglement entropy for a deformed planar entangling surface, finding  surprises at second order. 
}

\begin{document}
\maketitle
 
\section{Introduction}
The study of entanglement entropy is a rapidly developing field with applications in a broad range of contexts \cite{Bomb86, Sred93, SusUgl94, Callan94, CalCardy04,RefMoore04,KitPres05,RT, Faulkner:2013ica, Swingle:2014uza}. The utility of entanglement entropy, as well as the simplicity of its realization in holography \cite{RT}, suggests it has deep underlying structure hidden within it. It is therefore desirable to have a field-theoretic understanding of entanglement entropy based on first principles. In particular, one would like to calculate the dependence of entanglement entropy on the couplings of the theory as well as on the shape of the entangling surface and the background geometry. 

Entanglement entropy is given by the von Neumann entropy of the reduced density matrix for a subregion. The vacuum of a quantum field theory (QFT),  and in turn the reduced density matrix, can be defined by a Euclidean path integral. This suggests that one can study the change in the reduced density matrix induced by a deformation of the theory through a perturbative expansion of the action within the path integral. One can then find the resulting change in the entanglement entropy though a perturbative expansion of the von Neumman entropy of the density matrix. Through a proper choice of coordinates, one can treat geometric perturbations in a similar manner. In \cite{RS1, RS2, RS3} such an approach was initiated, giving expressions for entanglement entropy in terms of correlation functions. It should be emphasized that these correlation functions are evaluated on the original Euclidean manifold. Unlike the replica-trick method \cite{Callan94,Fursaev:1995ef,Lewkowycz:2013nqa,Fur13}, this computation avoids the technical challenges of computing correlation function on a replicated manifold. 

In this paper we continue the approach of \cite{RS1, RS2, RS3}\footnote{For a generalization of this approach to Renyi entropies, see \cite{Lewkowycz:2014jia}. For applications, see \cite{conifold,Allais:2014ata,Mezei:2014zla}.}, finding the change in the entanglement entropy to second order. Throughout we perturb around a CFT and a planar entangling surface in flat space.  One of our main new technical results is a calculation of the universal part of entanglement entropy for a general CFT perturbed by a relevant operator, up to second order in the coupling. This result may help in better understanding c-theorems and RG flows \cite{Zamolodchikov:1986gt,Cardy:1988cwa, Komargodski:2011vj} in the context of entanglement entropy \cite{Myers:2010xs,Myers:2010tj,Jafferis:2011zi,Klebanov:2011gs,Liu:2012eea,Casini12,Solodukhin:2013yha}, see also \cite{Sachdev, Klebanov:2012yf, Hertzberg:2012mn, Lewkowycz:2012qr, Agon:2013iva,Solodukhin:2014dva, Giombi:2014xxa}.

In Sec.~\ref{sec:relevant} we consider a theory deformed by a relevant operator, finding the dependence of entanglement entropy on the coupling of the operator. In Sec.~\ref{sec:pert} we review how to compute entanglement entropy perturbatively in the coupling, leading to an expression for the entanglement entropy in terms of correlation functions involving the stress-tensor and the relevant operator. In Sec.~\ref{sec:scalar} we warm up with a simple example: the entanglement entropy for a free massive scalar, where the mass term is treated as a relevant perturbation of the massless theory. Then in Sec.~\ref{sec:rel} we consider a general CFT deformed by a relevant operator $\mathcal{O}$ of dimension $\Delta$, with a small coupling $\lambda$. An explicit expression for the universal entanglement entropy is found in terms of $\Delta$ and $d$, up to second order in the coupling $\lambda$.

In Sec.~\ref{sec:geom} we consider the entanglement entropy for a CFT for a deformed entangling surface and weakly curved background. We first review how through an appropriate choice of coordinates adopted to the entangling surface (essentially a generalization of Gaussian normal coordinates), one can package both the change in the shape of the surface, and the background curvatures, into a metric perturbation $h_{\mu \nu}$. Thus, one can regard geometric perturbations as a change in the action of the field theory. The perturbative expansion then proceeds in a largely similar manner as in the context of relevant perturbations. At first order in the metric deformation, perturbative results~\cite{RS1} are in agreement with results in the literature \cite{solo,Fur13}. At second order, the situation is more subtle. Rather than doing the explicit second order calculation, we analyze on general grounds the structure of the possible result. 
Specifically, we consider all possible contractions of $h_{\mu \nu}$ consistent with symmetries, and make some assumptions on the form of the contact terms that will be considered. Demanding the terms in the perturbative expansion sum into a quantity consistent with reparameterization invariance along the entangling surface is sufficient to demonstrate that it is not possible for the result to fully agree with Solodukhin's expression \cite{solo} for the entanglement entropy for a $4$ dimensional CFT for a general entangling surface and general background. In particular, the relative coefficients of several of the curvature terms can not agree. The result of \cite{solo} was obtained through a combination of general arguments and holography, as well as squashed cone techniques~\cite{Lewkowycz:2013nqa,Fur13}, and checked holographically in \cite{Hung:2011xb}. Our faith in Solodukhin's expression leads us to believe that the only reconciliation is either a novel type of contact term that contributes, or the presence of a `non-perturbative' boundary term which gives an additional contribution to the entanglement entropy. Speculations to this effect can be found in the Discussion \ref{sec:discussion}.

\section{Relevant Perturbations} \label{sec:relevant}
The goal of this section is to find the dependence of entanglement entropy, $S$, on the coupling $\lambda$ of some relevant operator $\mathcal{O}$ (see Eq.~\ref{eq:relcorr}). For a planar entangling surface in flat space we explicitly carry out the relevant integrals of the correlation functions, finding $S$ to second order in $\lambda$ (see Eq.~\ref{main}).

\subsection{A perturbative expansion} \label{sec:pert}

Let us consider a general QFT that lives on a  $d$-dimensional Euclidean manifold $\M$ equipped with a Riemannian metric $g_{\mu\nu}$. The action of the field theory is given by $I(\phi, g_{\mu\nu})$, where $\phi$ collectively denotes all the QFT fields. For simplicity we assume that the system resides in the vacuum state $|0\rangle$. 

Consider now an arbitrary subregion $V$ of the manifold $\mathcal{M}$. The reduced density matrix for this region is obtained by tracing out the degrees of freedom associated with $\overline{V}\,$-$\,$the complement of $V$,
\be \label{eq:rho}
\rho = \text{Tr}_{\overline{V}} |0\rangle \langle 0| \equiv e^{-K} ~.
\ee
The right-hand side of (\ref{eq:rho}) serves as the definition of the modular Hamiltonian $K$. The entanglement entropy is defined as the von Neumann entropy of the reduced density matrix, 
\be \label{eq:Sdef}
S = -\text{Tr}_V\,(\rho\log\rho) = \langle 0| K |0\rangle~.
\ee
\begin{figure}[tbp] 
\centering
\subfigure[]{
	\includegraphics[width=2in]{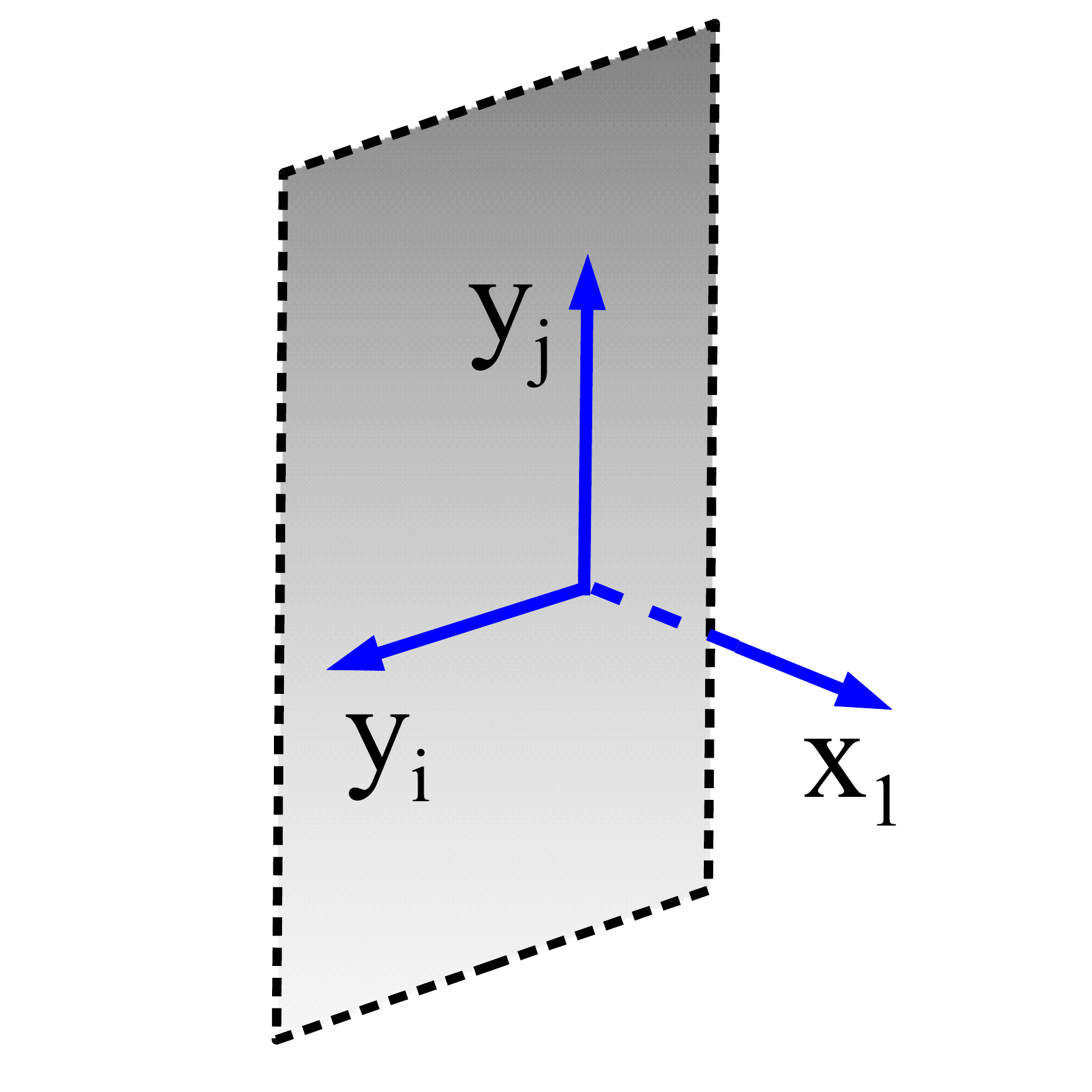}
	}
		\hspace{.2in}
		\subfigure[]{
	\includegraphics[width=2in]{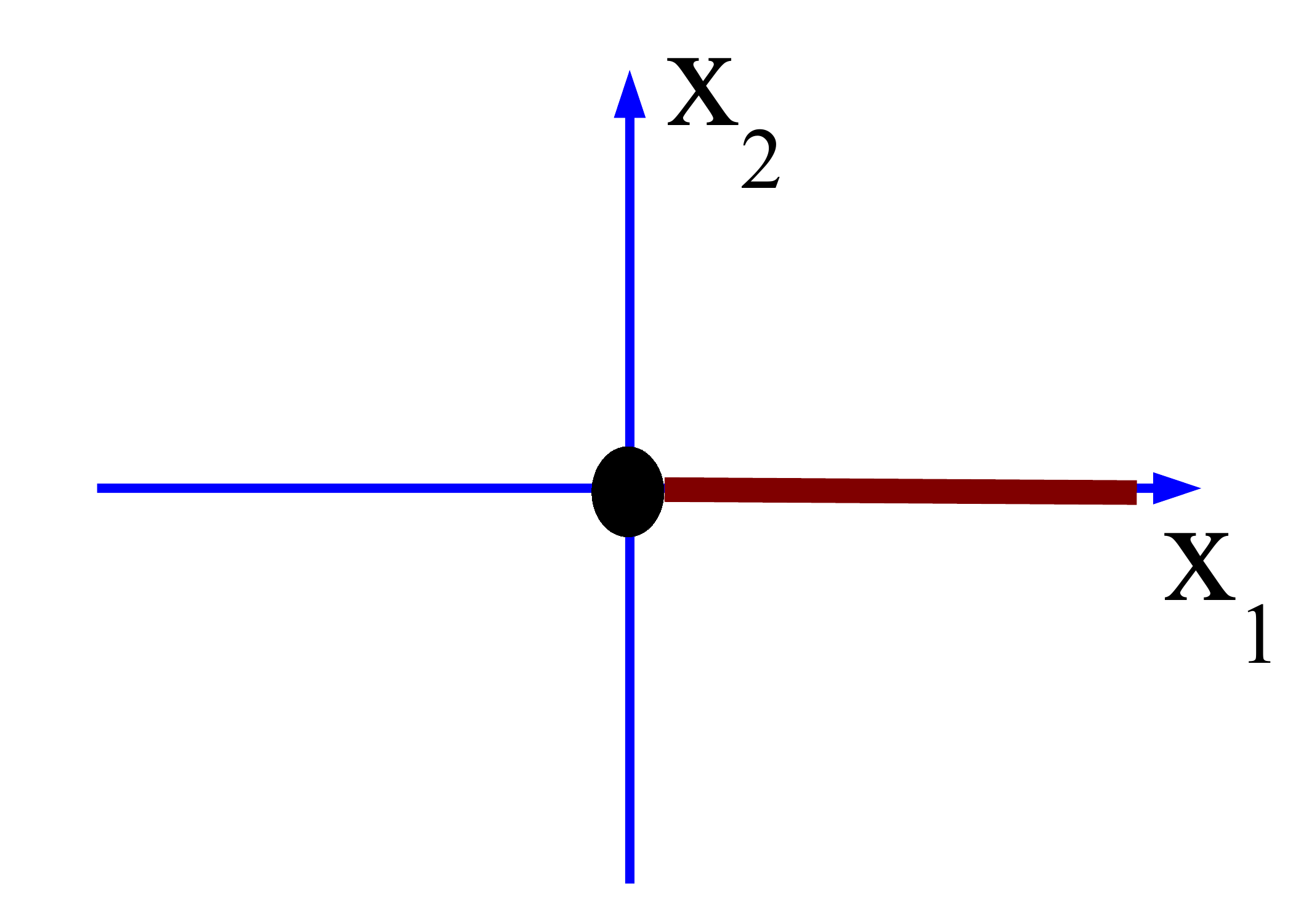}
	}
\caption{(a) An entangling surface $\Sigma$ that is a plane. We use coordinates $x_{\mu} = (x_a,y_i)$, with $x_a$ transverse to the plane and $y_i$ along the plane. (b) The transverse space to the plane.} 
\label{fig:plane}
\end{figure}

The vev on the right-hand side can be regarded as a Euclidean path integral over the entire manifold with insertion of $K$ along the cut $\mathcal{C}$ through the subregion $V$. This cut $\mathcal{C}$ corresponds to a subregion of some constant time slice where the modular Hamiltonian is defined. 

We are interested in finding the dependence of $S$ on $\lambda$. We therefore take derivatives of (\ref{eq:Sdef}) with respect to the coupling $\lambda$ associated with $\mathcal{O}$,
\be
 {\del S\over \del\lambda}=-\langle \mathcal{O}  K \rangle + \langle {\del K\over \del\lambda} \rangle~,
 \label{first}
\ee
where $\langle\cdots\rangle$ denotes the vev, $\langle K \mathcal{O}(x)\rangle$ is a connected correlation function, and $\mathcal{O}$ stands for the integral of the local operator $\mathcal{O}(x)$ over the entire manifold. The second term on the right-hand side vanishes since, by assumption, $K$ is such that the density matrix is normalized, 
\be
 0={\del\over \del\lambda}\text{Tr} (e^{-K}) = - \text{Tr} \big( {\del K \over \del\lambda} e^{- K} \big) = -\langle {\del K \over \del\lambda} \rangle~.
 \labell{norm}
\ee
Hence, we have the entanglement flow equation \cite{RS2}
\be
 {\del S\over \del\lambda}=-\langle \mathcal{O}  K \rangle ~.
 \labell{relflow}
\ee
Now taking a second derivative with respect to $\lambda$ yields,
\be
 {\del^2 S\over \del\lambda^2}=-{\del\over\del\lambda}\langle \mathcal{O}  K \rangle = \langle \mathcal{O}\mathcal{O}  K \rangle 
 -\langle \mathcal{O} \, {\del K\over \del\lambda}  \rangle~.
\ee
Substituting these results into a Taylor expansion of $S$ we obtain,
\be
 \delta S=-\langle \mathcal{O}  K \rangle \, \delta\lambda+{\delta\lambda^2\over 2} \Big( \langle \mathcal{O}\mathcal{O}  K \rangle 
 -\langle \mathcal{O} \, {\del K\over \del\lambda}  \rangle\Big) +\ldots ~.
 \label{2nd}
\ee
The above expression is completely general since no assumption has been made about the unperturbed theory, or the geometry of the background and the entangling surface. Of course, in general the modular Hamiltonian $K$ is unknown. 

An exceptional case is that of a planar entangling surface embedded in flat space, for which the modular Hamiltonian is proportional to the Rindler Hamiltonian, 
\begin{equation} \label{eq:Kplane2}
K = -2\pi\int_{\Sigma} \int_0^\infty dx_1\, x_1\, T_{22} ~,
\end{equation}
where $\Sigma$ is the entangling surface, $x_1, x_2$ are orthogonal to $\Sigma$ (see Fig.\ref{fig:plane}), and the energy-momentum tensor is defined by varying the Euclidean action, $I$, with respect to the background metric,
\be
 T_{\mu\nu}(x)=-{2\over \sqrt{g}} {\delta I\over \delta g_{\mu\nu}(x)}~.
 \labell{foot8}
\ee 
Since the dependence of the stress-tensor on $\lambda$ is of the form,
\be \label{eq:Tmunu}
 T_{\mu\nu}(x)=T_{\mu\nu}^{0}(x)-\delta_{\mu\nu}\,\lambda\,\mathcal{O}(x)~,
\ee
where $T_{\mu \nu}^{0}$ is the energy-momentum tensor of the theory with $\lambda =0$, we have that for a planar entangling surface, the derivative of the modular Hamiltonian in \reef{2nd} can be replaced with $\mathcal{O}$.\footnote{In Appendix \ref{sec:general} we give some evidence that such a replacement might be true more generally.} Thus we have, \footnote{Due to symmetry in the transverse plane, some codimension-$1$ integrals can be replaced by full $d$-dimensional integrals \cite{RS3}.}~\footnote{A slightly different derivation of (\ref{eq:relcorr}) consists of starting with the path integral definition of $\rho$, perturbatively expanding the action, and inserting $\rho_0 + \delta \rho$ into a perturbative expansion of the definition of the von Neumann entropy \cite{RS1, RS2}. There is also a slight variation of the above derivation of (\ref{eq:relcorr}) in which one  at the outset inserts the definition of the modular Hamiltonian (\ref{eq:Kplane2}) into the flow equation (\ref{relflow}). Then the expansion of the right-hand side is the familiar kind of field theory expansion of the correlator $\langle T \mathcal{O}\rangle$ for one theory in terms of correlation functions of a theory with different couplings \cite{RS3}.}

\be
 \delta S=-\langle \mathcal{O}  K \rangle \, \delta\lambda+  {(\delta \lambda)^2\over 2} \Big( \langle  \mathcal{ O}\mathcal{O} K\rangle -  \langle \mathcal{ O}\mathcal{ O}\rangle\Big)+\ldots~.
 \label{eq:relcorr}
\ee
The rest of the section will focus on explicitly evaluating (\ref{eq:relcorr}).

\subsection{Warmup: free massive scalar} \label{sec:scalar}
A simple context in which to test (\ref{eq:relcorr}) is that of the entanglement entropy for a massive free scalar field $\phi$. Regarding the mass term as a deformation of the massless theory, we have $\mathcal{O} = \phi^2$ with coupling $\delta \lambda = m^2/2$. In even spacetime dimensions, the entanglement entropy contains a logarithmic divergence which does not depend on the details of the regularization scheme, and  is therefore regarded as universal. For the  scalar field, this universal part of entanglement entropy takes the form \cite{Hertzberg:2010uv,Huerta:2011qi,Lewkowycz:2012qr},
\be \label{eq:SscalarF}
  S=   {(-)^{d\over 2}\over 6 (4\pi)^{d-2\over 2}\Gamma(d/2)} 
  \, m^{d-2}\A_\Sigma \, \log(m\delta)  ~,
\ee
where $\A_\Sigma$ is the area of the entangling surface and $d$ is the spacetime dimension.
In $4$ dimensions, $S \sim m^2 \A_\Sigma\, \log(m\delta)$ and can therefore be found from the linear term in (\ref{eq:relcorr}), see \cite{RS3}. Here, our interest is to test the quadratic term in (\ref{eq:relcorr}). Since in $6$ dimensions, $S \sim m^4 \A_\Sigma\, \log(m\delta)$, we can therefore use (\ref{eq:relcorr}) to recover (\ref{eq:SscalarF}) in $6$ dimensions.

Evaluating (\ref{eq:SscalarF}) amounts to simply preforming some integrals of correlation functions of a free massless scalar field theory. For the canonically normalized scalar field, the two-point functions are,
\begin{equation} \label{eq:2point}
\langle \phi(x) \phi(0) \rangle = \frac{1}{(d-2) S_{d}} \frac{1}{x^{d-2}} \quad \Rightarrow \quad \langle \mathcal{O}(x)\mathcal{O}(0)\rangle = \frac{2}{(d-2)^2 S_{d}^2} \frac{1}{x^{2(d-2)}}~,
\end{equation}
where $S_d = 2 \pi^{d/2}/ \Gamma(d/2)$ is the solid angle.  The canonical energy-momentum tensor is 
\begin{equation} \label{eq:Tscalar}
T_{\mu \nu}^{0} = \partial_{\mu} \phi \partial_{\nu} \phi - \frac{1}{2} \delta_{\mu \nu} (\partial \phi)^2~.
\end{equation}
Now using (\ref{eq:2point}) and  (\ref{eq:Tscalar}), it follows through  Wick contractions that 
\begin{equation} \label{eq:TO}
\langle T_{\mu\nu}^{0}(\bar{x})\, \phi^2(x)\rangle = \frac{2(x_\mu - \bar{x}_\mu)(x_\nu - \bar{x}_\nu) - \delta_{\mu\nu}(x-\bar{x})^2 }{S_d^2\, (x-\bar{x})^{2d}}~.
\end{equation}
Similarly, the three-point function is 
\begin{multline} \label{eq:TOO}
\langle T_{\mu\nu}(x)\, \phi^2(\bar x)\, \phi^2(z) \rangle =\\
 \frac{4}{(d-2)\, S_d^3} \frac{(x_\mu - \bar x_\mu)(x_\nu - z_\nu)+(x_\nu - \bar x_\nu)(x_\mu - z_\mu)
 - \delta_{\mu\nu}(x_{\rho} - \bar x_{\rho})(x^{\rho} - z^{\rho})}{(x - \bar x)^d\, (x- z)^d\, (\bar x-z)^{d-2}}~.
\end{multline}
The two terms in (\ref{eq:relcorr}) that we need to evaluate  are $\langle K \mathcal{O} \mathcal{O}\rangle$ and $\langle \mathcal O \mathcal{O} \rangle$. We start with $\langle \mathcal O \mathcal{O} \rangle$,
\be \label{eq:OO}
 \langle \mathcal{  O}\mathcal{  O}\rangle \equiv \int d^dx \int d^dz\, \langle \mathcal{  O}(x)\mathcal{  O}(z)\rangle~.
\ee
The integrals on the right-hand side exhibit both UV and IR divergences. Indeed, using the translational symmetry of the measure and the two-point function (\ref{eq:2point}) yields,
\be
\langle \mathcal{  O}\mathcal{  O}\rangle = \frac{2}{(d-2)^2 S_{d}^2} \int d^dx \int d^dz \frac{1}{x^{2(d-2)}}~.
\label{OOtrans}
\ee
If (\ref{OOtrans}) were true, then the final answer in 6 dimensions would be given by a product of an IR divergent volume and a UV divergent integral. However, given the divergent behavior of the initial integral \reef{eq:OO}, it is apparent that the manipulations leading to \reef{OOtrans} are too na\"{\i}ve. 

To disentangle the divergences, we first exploit the rotational symmetry in the transverse space to the entangling surface to rewrite (\ref{eq:OO}) as \cite{RS3}
\be \label{eq:OO1}
\langle \mathcal{O} \mathcal{O} \rangle =  - \frac{2 \, (2\pi)^2}{(d-2)^2 S_d^2}\, \int_{-\infty}^{0}\, d\bar{x}_1\, \bar{x}_1\, \int d^{d-2} \bar{y} 
\int_{0}^{\infty}\, d x_1\, x_1 \int d^{d-2} y \frac{1}{(\bar{x} - x)^{2(d-2)}} ~,
\ee
where our notation is $x^{\mu} = (x^a,y^i)$, with $\{y^i\}_{i=1}^{d-2}$ along the entangling surface and $\{x^a\}_{a=1}^2$ transverse to the surface (and similarly for $\bar x^a$ and $\bar y^i$).
Substituting $d=6$ and carrying out the convergent integrals over $\bar y$, $y$ and $\bar x_1$, we obtain \footnote{Note that after integrating over $\bar y$, the integrand becomes $y$-independent. In particular, $\A_\Sigma$ denotes the integral over $y$ which represents the (infinite) area of the entangling plane.}
\be \label{eq:OO2}
\langle \mathcal{O} \mathcal{O} \rangle = \frac{ \A_\Sigma}{72 \pi^2} \int_\delta^{m^{-1}} {d x_1\over x_1}
= - \frac{\A_\Sigma}{72 \pi^2}  \log(m \delta) ~,
\ee
where we introduced a UV cut-off $\delta$ and an IR cut-off $m^{-1}$ to regularize divergences of the integral. 
We now turn to the $\langle K \mathcal{O} \mathcal{O}\rangle$ term in (\ref{eq:relcorr}), which can be written as,
\be \label{eq:KOO1}
\langle K \mathcal{O} \mathcal{O} \rangle =(2\pi)^2 \int d^{d-2} y \int_0^\infty dx_1 \, x_1\int d^{d-2} \bar y  \int_{-\infty}^0 d\bar x_1\, \bar x_1  \int d^d z \,
 \langle T_{22}(x)\, \phi^2(\bar{x})\, \phi^2(z) \rangle ~.
\ee
The integrals in (\ref{eq:KOO1}) can be carried out through the use of (\ref{eq:TOO}), and are performed in Appendix \ref{sec:appendix}, yielding in $d=6$,
\be \label{eq:KOO2}
\langle K \mathcal{O} \mathcal{O} \rangle =  - \frac{\A_{\Sigma}}{18 \pi^2}  \log(m \delta) ~.
\ee
Thus, combining (\ref{eq:relcorr}) with (\ref{eq:OO2}) and (\ref{eq:KOO2}), we finally obtain
\be
\delta S =  - \frac{m^4 \A_{\Sigma}}{192 \pi^2} \,  \log(m \delta)~,
\ee
matching the known result (\ref{eq:SscalarF}).

\subsection{Perturbed CFT} \label{sec:rel}
We now turn to evaluating (\ref{eq:relcorr}) for the general case of a CFT deformed by a relevant operator $\mathcal{O}$ of scaling dimension $\Delta$,
\be
I = I_0 + \lambda \int{d^d x\, \mathcal{O}}(x)~,
\ee
where $I_0$ is the CFT action. 

For a CFT, the correlator $\langle T_{\mu \nu} \mathcal{O}\rangle$ vanishes. Correspondingly, $\langle K \mathcal{O}\rangle$ vanishes, as $K \sim T_{\mu \nu}$ for a planar entangling surface (\ref{eq:Kplane2}). Therefore, the first nonvanishing contribution will occur at second order in $\lambda$. The CFT correlation functions which will be relevant are \cite{Osborn:1993cr, Erdmenger:1996yc}\footnote{In equations (\ref{2pO}), (\ref{TOO}) and (\ref{defs}) we are using the notation of \cite{Osborn:1993cr} in which the subscript on $x$ denotes different points, as opposed to our notation in which it refers to different components of a vector field $x$.}
\be
 \langle \hO (x_2) \hO(x_3) \rangle = {N\over (x_2-x_3)^{2\Delta}}
 \labell{2pO}
\ee
and
\be
 \langle T_{\mu\nu}(x_1) \hO (x_2) \hO(x_3) \rangle ={1\over x_{12}^d \, x_{23}^{2\Delta-d} \, x_{31}^d} \, t_{\mu\nu} \big( \hat X_{23}\big)~,
 \labell{TOO}
\ee
where\footnote{The sign flip in the definition of $`a$' relative to \cite{Osborn:1993cr} is a result  of a  corresponding sign flip in the definition of the energy-momentum tensor.}
\bea
 t_{\mu\nu}(\hat X) &=&a\ \( \hat X_\mu \hat X_{\nu} - {1\over d} \delta_{\mu\nu}\)~, 
\quad
 X_{23}={x_{21}\over x_{21}^2 }- {x_{31} \over x_{31}^2} ~, 
 \quad 
 X_{23}^2={x_{23}^2\over x_{21}^2 x_{31}^2}~, 
 \non
  a&=&{d\Delta N\over (d-1) S_d} ~.
 \labell{defs}
\eea 
Equipped with these correlators, we turn to  explicitly evaluating \reef{eq:relcorr}.

We start with the two-point function of $\mathcal{   O}$
\be
 \langle \mathcal{   O}\mathcal{   O}\rangle=\int d^dx \int d^dz \langle \mathcal{  O}(x)\mathcal{  O}(z)\rangle~.
\ee
The integrals should be carefully treated due to the IR and UV divergences that may interfere. Hence, we first exploit the rotational symmetry inherent to the entangling surface to rewrite the above correlator as \cite{RS3}
\be
 \langle \mathcal{  O}\mathcal{  O}\rangle=-(2\pi)^2 \int d^{d-2} \bar y  \int_{-\infty}^0 d\bar x_1\, \bar x_1 \int d^{d-2} y \int_0^\infty dx_1 \, x_1
 {N\over \big( (y-\bar y)^2+(x_1-\bar x_1)^2\big)^\Delta}~.
\ee
Next, we carry out the integrals over $\bar y$, $y$ and $\bar x_1$. The final result takes the form
\be
  \langle \mathcal{  O}\mathcal{  O}\rangle={N\, \pi^{d/2+1}\over \Delta-(d-1)/2} \, {\Gamma(\Delta-d/2)\over \Gamma(\Delta)}
  \A_\Sigma \int_\delta^\ell d x_1 ( x_1)^{d-2\Delta+1}~,
  \labell{OO}
\ee
where we introduced an IR cut-off $\ell=\lambda^{1\over \Delta-d}$.

Furthermore,
\be
 \langle   K \mathcal{  O}\mathcal{  O}\rangle=(2\pi)^2 \int d^{d-2} y \int_0^\infty dx_1 \, x_1\int d^{d-2} \bar y  \int_{-\infty}^0 d\bar x_1\, \bar x_1  \int d^d z
 \, \langle T_{22}(x) \mathcal{O}(\bar x) \mathcal{O}(z) \rangle~.
 \labell{KOO}
\ee
The integrals in the above expression can be evaluated  using \reef{TOO} and \reef{defs}; the details are in Appendix \ref{sec:appendix} . The result is 
\be
 \langle   K \mathcal{  O}\mathcal{  O}\rangle={4\,a\,\pi^{d+1}\over d (d-2\Delta)(d-2\Delta-1)}\,{\Gamma(\Delta-d/2)\over\Gamma(\Delta)\Gamma(d/2)}
 \A_{\Sigma}\int_\delta^\ell dx_1 x_1^{d-2\Delta+1}~.
\ee
Substituting this result and \reef{OO} into \reef{eq:relcorr} yields,
\be
 \delta S= \lambda^2 { \pi^{d/2+1}\over (d-2\Delta-1)} \, {\Gamma(\Delta-d/2)\over \Gamma(\Delta)}
 \Big( {a \, S_d\over d (d-2\Delta)}+ N\Big)
 \A_{\Sigma}\int_\delta^\ell dx_1 x_1^{d-2\Delta+1}+\mathcal{O}(\lambda^3)~.
\ee
The universal divergence emerges at second order in the relevant coupling $\lambda$ if and only if the scaling dimension of $\mathcal{O}(x)$ is $\Delta=(d+2)/2$. In this case
\be
 \delta S=  N \lambda^2 \, {d-2 \, \over 4(d-1) } \, {\pi^{d+2\over 2}\over \Gamma\({d+2\over 2}\)}
 \A_{\Sigma}\log(\delta/\ell)~, \quad \Delta={d+2\over 2} ~.
 \labell{main}
\ee

Eq.~\reef{main} is one of our main results, expressing the universal entanglement entropy  that arises from a relevant deformation of a CFT, for a planar entangling surface.
A few comments:
\begin{itemize}
\item The result is valid for $\textit{any}$ CFT deformed by a relevant operator. Remarkably, its form is independent of what the CFT is. This property is inherited from the universality of the 3-point function $\langle T_{\mu \nu} \mathcal{O} \mathcal{O}\rangle$. Furthermore, (\ref{main}) is valid for both a weakly and strongly coupled CFT.
\item Eq.~(\ref{main}) reproduces the known results for the special case of massive free fields (see Appendix \ref{appendixB}).
\item Eq.~(\ref{main}) reproduces the holographic calculation performed in \cite{Hung:2011ta}, see also \cite{Liu:2012eea}. From the 2-point function obtained via AdS/CFT \cite{Klebanov:1999tb} one has \be
N =  2\, {\Delta-d/2 \over \pi^{d/2}}  {\Gamma(\Delta) \over \Gamma(\Delta-d/2)} \, \eta\,L_{\mt{AdS}}^{d-1}~,
\ee
where $\eta$ is the normalization of the action for the bulk scalar field which couples to $\mathcal{O}$. Choosing $\eta = (2 l_{\mt{Planck}}^{d-1})^{-1}$ as in \cite{Hung:2011ta}, we find \reef{main} matches (3.26) of \cite{Hung:2011ta}.\footnote{We thank Mark Mezei for discussion on this point.}

\item Eq.~\reef{main} is more general than may appear. Although \reef{main} was derived for a specific geometry,  it will in fact be a contribution to the entanglement entropy for a slightly perturbed CFT for any background and any entangling surface, with $\mathcal{A}_{\Sigma}$ being the area of the particular entangling surface. This is simply a manifestation of the fact that any entangling surface and any background look flat in a sufficiently small neighborhood of the surface. Of course, for a non-planar entangling surface in a curved background, there will be additional contributions to the entanglement entropy that involve the curvatures. 
\item Eq.~\reef{main} was derived for a relevant perturbation of a CFT. Nothing in the formalism requires deforming around a CFT. One can repeat the computation, deforming around any theory, provided one knows the low point correlation functions, (\ref{2pO}) and (\ref{TOO}). 
\end{itemize}

\section{Geometric Perturbations} \label{sec:geom}

In Sec.~\ref{sec:relevant} we studied the universal entanglement entropy for a CFT deformed by a relevant operator. In this section, the focus is on geometric deformations, resulting from either a slightly curved background, or a slight deformation in the shape of the entangling surface. Through a proper choice of coordinates adopted to the entangling surface, both deformations can be regarded as a perturbation, $h_{\mu\nu}$, of the flat Euclidean metric \cite{RS1}. At linear order, the change in the action is $I = I_0 - {1\over 2}\int T^{\mu \nu} h_{\mu \nu}$. As a result, in many respects geometric and relevant deformations are similar. There are, however, important differences: the action depends non-linearly on $h_{\mu\nu}$, which in turn is a non-constant function on $\M$. 

The first order calculation for geometric perturbations was carried out in \cite{RS1}. In this section, we find on general grounds the possible form of the second order contribution, and compare it with \cite{solo}. The explicit computation of the second order contribution is relegated to Appendix \ref{AppendixC}.

\subsection{A perturbative expansion}
Let us consider a given entangling surface and background, $(\Sigma, \M)$, that undergoes a slight deformation. As argued in \cite{RS1}  one can find a foliation of the space such that the details of any geometric perturbation are encoded in the coefficients of a Taylor expansion of the metric in the vicinity of $\Sigma$. In particular, any small deformation of the geometry  induces a  small change in the coefficients of such an expansion. Therefore, splitting the metric into a background part $g_{\mu\nu}$ and a small perturbation $h_{\mu\nu}$ is well-defined. 

Moreover, since the UV divergences of entanglement entropy are local, only being sensitive to the quantum fluctuations in the vicinity of $\Sigma$, such an expansion of the metric is enough to evaluate the variation of the universal entanglement entropy. In our case, the unperturbed geometry corresponds to a planar $\Sigma$ in flat space. Hence, $g_{\mu\nu}=\delta_{\mu\nu}$ and \cite{RS1}
\begin{eqnarray}
h_{ij} &=& \delta\gamma_{ij}+2 K_{\ha ij} \, x^\ha+  x^\ha x^{\hc}\big( \delta_{\ha\hc} A_iA_j+\R_{i\ha \hc j}|_\Sigma +K_{\hc \, i l} K_{\ha\,j}^{~l}  \big) + \mathcal{O}(x^3) ~, 
\non
h_{ab} &=& -\frac{1}{3} \mathcal{R}_{a c b d}|_\Sigma x^c x^d + \mathcal{O}(x^3) ~,
\label{h}
\\
h_{i c} &=& {1\over 2}\big(A_i+{1\over 3}x^\hb \veps^{\hd\he}\R_{i\hb\hd\he}\big|_{\Sigma} \big)x^a\veps_{\ha\hc}+ \mathcal{O}(x^3) ~ .
\nonumber
\end{eqnarray}
Here $\delta\gamma_{ij}$ and $K_{\ha ij}$ represent deformations in the induced metric on $\Sigma$ and the associated extrinsic curvatures, respectively, while $\R_{\mu\nu\rho\sigma}$ denotes the background Riemann tensor. The vector field $A_i$ lives on the surface, and is analogous to a Kaluza-Klein gauge field. This kind of foliation has recently been useful in the study of holographic computations \cite{Lewkowycz:2013nqa, Fur13, Dong:2013qoa, Camps:2013zua}. Note that $\delta\gamma_{ij}$, $K_{\ha ij}$, $\R_{\mu\nu\rho\sigma}$ and $A_i$ are non-constant tensors on $\Sigma$ that contain at most two derivatives of the background metric, whereas higher order terms in \reef{h} include at least three derivatives. To leading order, $\delta\gamma_{ij}$, $K_{\ha ij}$, $\R_{\mu\nu\rho\sigma}$ and $A_i$ are linear in a small parameter that characterizes a given geometric deformation. Therefore, to second order in the deformation, the change in the entanglement entropy is given by,
\be
 \delta S=\int d^dx  \frac{\delta S}{\delta g_{\mu \nu}(x)}h_{\mu\nu}(x)
 +{1\over 2}\int d^dx \int d^d\bar x \frac{\delta^2 S}{\delta g_{\al\bt}(x)\delta g_{\mu \nu}(\bar x)} h_{\al \bt}(x)h_{\mu\nu}(\bar x)
 +\ldots ~.
 \label{Staylor}
\ee

Our main interest in what follows is to find the universal (or logarithmic) divergence of entanglement entropy for a CFT in four space-time dimensions. Furthermore, since our goal is to try to recover \cite{solo}, we are interested in the contribution to entanglement entropy that is exclusively a local geometric combination of the background and extrinsic curvatures. As the entangling surface is two dimensional, by dimensional analysis, the only terms that can appear are two-derivative terms: the background curvature and quadratic combinations of the extrinsic curvatures. As a result, the terms explicitly presented in \reef{Staylor}, combined with \reef{h}, are enough to capture the structure of universal entanglement entropy in four space-time dimensions.

From a computational point of view, one starts with \reef{Staylor} and integrates out a two-dimensional transverse space, only keeping track of the logarithmically divergent contribution.
By construction, the linear term in \reef{Staylor} will generate a local contribution to the universal entanglement entropy, whereas the contribution of the second term is in general not local, as  the $h$'s are evaluated at different points. That said, a local contribution will emerge if the second variation of $S$ contains a delta function which identifies the arguments of the two $h$'s, and this is the form we will assume.\footnote{ This delta function is associated with a possible contact term which is inherent to correlation functions of operators with overlapping support. In Appendix \ref{AppendixC} such terms are explicitly found.}

\subsection{Geometric structures}
Let us consider the possible structures on the right-hand side of \reef{Staylor}. We have both rotational and translational symmetry along the entangling surface $\Sigma$, while the admissible tensors in the transverse space are $\delta_{ac}$ and $x^a$. Thus, the most general form of the first and second variations of  entanglement entropy are,
\bea
\frac{\delta S}{\delta g_{a i}(x)}&=&0~,
\quad
\frac{\delta S}{\delta g_{ij}(x)}=\alpha_1{\delta^{ij}\over r^4}~,
\quad
\frac{\delta S}{\delta g_{ac}(x)}= \al_2\, {\delta^{ac}\over r^4} + \al_3 \, {x^a x^c \over r^6} ~,
\non
\frac{\delta^2 S}{\delta g_{ij}(x)\delta g_{kl}(\bar x)} &=&\(\alpha_4\,{\delta^{ij}\delta^{kl} \over r^4}
+ \alpha_5\,{ \delta^{i(k}\delta^{l)j}\over r^4}\)\delta(x-\bar x)+\ldots~,
\label{varS2}\\
\frac{\delta^2 S}{\delta g_{ac}(x)\delta g_{ij}(\bar x)} &=&\( \al_6\, {\delta^{ac}\over r^4} + \al_7 \, {x^a x^c \over r^6} \)\, \delta^{ij}\,\delta(x-\bar x)+\ldots~,
\nonumber
\eea
where $\{\alpha_i\}_{i=1}^7$ are some constants, $r^2=x_1^2+x_2^2$ represents the radial distance in the transverse space, and in the last two equations we suppressed all terms which will not give a local contribution to the entanglement entropy. \footnote{\label{footc} In (\ref{varS2}) we have not included contact terms that involve derivatives of the delta function. On dimensional grounds, such terms must then be multiplied by powers of distance. Derivatives of the delta function multiplied by powers of $(x - \bar{x})$ will not lead to any new contributions. On the other hand, it is a logical possibility that there are terms having powers of $r$ multiplying derivatives of the delta function. However, these would be very nonstandard contact terms, in that they would depend not only on the relative separation between the two points, but also on the absolute distance ($r$) of the points. Certainly in the context in which the contact terms arise from the collision of stress-tensors inside of correlation functions, such terms are absent. Nevertheless, given that we will be unable to recover \cite{solo}, it may be worth thinking more about ways in which this assumption can be violated. }

Substituting these general expressions into \reef{Staylor} gives,
\bea
 \delta S&=&\int d^dx \, {\al_1\over r^4}\, \delta^{ij}h_{ij}+\int d^dx \(  \al_2\, {\delta^{ac}\over r^4} + \al_3 \, {x^a x^c \over r^6} \) h_{ac} 
 \labell{Staylor2}
 \\
  &+&{1\over 2}\int d^dx \Big({\al_4\over r^4} \, (\delta^{ij}h_{ij})^2  +{\al_5\over r^4} \, h_{ij} h^{ij}\Big) 
  +\int d^dx \( \al_6\, {\delta^{ac}\over r^4} + \al_7 \, {x^a x^c \over r^6} \)\, h_{ac}\,\delta^{ij}h_{ij}+\ldots ~,
 \nonumber
\eea
where we suppressed $\mathcal{O}(h_{ac}^2)$ and $\mathcal{O}(h_{ai}^2)$ terms since they do not contribute to the universal entanglement entropy. \footnote{Indeed, according to \reef{h} the $\mathcal{O}(h_{ac}^2)$ term has four derivatives and thus cannot contribute to the logarithmic divergence in four dimensions, whereas  $\mathcal{O}(h_{ai}^2)$ is quadratic in the gauge field $A_i$, and hence its net contribution must vanish.} Using \reef{h}  we get,
\begin{eqnarray}
\delta^{ij}h_{ij}&=& \delta\gamma + 2 K_{\ha} \, x^\ha+  x^\ha x^{\hc}\big( \delta_{\ha\hc} A_iA^i+\delta^{ij}\R_{i\ha \hc j}|_\Sigma +K_{\hc \, ij} K_{\ha}^{~ ij}  \big) + \mathcal{O}(x^3) ~, 
\non
(\delta^{ij}h_{ij})^2&=&\delta\gamma^2 + 4 \, \delta\gamma \, K_{\ha} \, x^\ha + 4 K_{\ha} \, K_{\hc} \, x^\ha x^\hc
+ 2 \, x^\ha x^{\hc} \, \delta\gamma \, \delta^{ij}\R_{i\ha \hc j}|_\Sigma + \mathcal{O}(x^3)~,
\label{hh}
\\
h^{ij}h_{ij}&=&\delta\gamma^{ij}\delta\gamma_{ij} + 4 \, \delta\gamma^{ij} \, K_{\ha \, ij} \, x^\ha + 4 K_{\hc \, ij} K_{\ha}^{~ ij} \, x^\ha x^\hc
+2 \, x^\ha x^{\hc} \, \delta\gamma^{ij} \, \R_{i\ha \hc j}|_\Sigma + \mathcal{O}(x^3)~,
\nonumber
\end{eqnarray}
where $\delta\gamma=\delta^{ij}\delta\gamma_{ij}$, and we have suppressed terms, such as $\delta\gamma^{ij}A_iA_j$, that are cubic in the small parameter of deformation.

As mentioned earlier, the $A_iA^i$ term appearing in \reef{hh} is a gauge term and can not contribute to $\delta S$: it must be canceled by a similar contribution from the $\mathcal{O}(h_{ai}h^{ai})$ term in \reef{Staylor}. Terms linear in the extrinsic curvatures $K_{a\,ij}$ are also irrelevant: they have only one derivative while dimensional analysis requires two-derivative terms.\footnote{From the computational point of view, these terms vanish since the integrand in \reef{Staylor2} is odd for these terms. Furthermore, extrinsic curvature is sensitive to the orientation of the surface while entanglement entropy is certainly orientation independent.}  Hence, combining \reef{Staylor2} and \reef{hh}, we get the general structure for the (local) universal entanglement entropy in four dimensions\footnote{$\al_3$ and $\al_7$ do not contribute since $R_{abcd}x^ax^bx^cx^d=0$.}
\bea
 \delta S&=&  \pi \, \al_1 \int d^2y \big(\delta^{ij}\delta^{ac}\R_{i\ha \hc j} +K_{ ij}^{\ha} K_{\ha}^{ij} \big) \log(\ell/\delta)
 -{\pi\, \al_2\over 3} \int d^2y \,\delta^{\ha\hc} \, \delta^{bd}\R_{a b c d}  \log(\ell/\delta)
 \non
 &+&\pi\,\al_4 \int d^2y \big(2  K^{\ha} \, K_{\ha} +  \delta\gamma \, \delta^{ij}\delta^{\ha \hc}\, \R_{i\ha \hc j}  \big) \log(\ell/\delta)
  \label{Suniv0}
 \\
 &+&\pi\,\al_5 \int d^2y \big( 2 K_{ ij}^{\ha} K_{\ha}^{~ ij}  +  \delta\gamma^{ij} \, \delta^{\ha\hc} \R_{i\ha \hc j}  \big) \log(\ell/\delta) 
   -{\pi\, \al_6\over 3} \int d^2y \,\delta\gamma\,\delta^{\ha\hc} \, \delta^{bd}\R_{a b c d}  \log(\ell/\delta)
 ~,
 \nonumber
\eea
where $\delta$ is the UV cut-off and we used
\bea
 \int d^2x  \, {x^a x^c \over r^4}  =  \pi \delta^{ac} \, \log(\ell/\delta) ~ ,
\eea
where $\ell$ is a characteristic scale of the deformed geometry.

Now, the explicit first order calculation \cite{RS1} found that $\al_2=-3\al_1=-c/(2\pi^2)$, with $c$ being the central charge of a CFT defined by the trace anomaly,
\be
 \langle T^\mu_\mu\rangle = {c\over 16\pi^2}\int_\M C_{\mu\nu\rho\sigma}C^{\mu\nu\rho\sigma} - {a\over 16\pi^2}\int_\M E_4~,
 \label{trace}
\ee
where $C_{\mu\nu\rho\sigma}$ is the Weyl tensor and $E_4$ is the Euler density in four dimensions. To fix the remaining coefficients $\al_4,\,\al_5$ and $\al_6$, we note that the terms in (\ref{Suniv}) should combine into an expression that is manifestly invariant under diffeomorphisms restricted to the entangling surface. For instance, the $\alpha_2$ and $\alpha_6$ terms should combine into an expression of the form $\int d^2 y \sqrt{\gamma} \delta^{ac} \delta^{bd} \R_{a b c d}  \log(\ell/\delta)$, and this fixes $\alpha_6 =\alpha_2/2$. The coefficients $\alpha_4$ and $\alpha_5$ are fixed similarly, 
\be \label{eq:cons}
\al_4={\al_1 \over 2}~,\quad \al_5=-\al_1~,\quad \al_6={\al_2\over 2}~.
\ee
Substituting into \reef{Suniv0} yields,
\bea
 \delta S= {c\over 6\pi} \int d^2y \sqrt{\gamma} \big(\gamma^{ij}\delta^{ac}\R_{i\ha \hc j} +\delta^{\ha\hc} \, \delta^{bd}\R_{a b c d}+ K_a K^a - K_{ ij}^{\ha} 
 K_{\ha}^{ij} \big) \log(\ell/\delta) 
\eea
where $\sqrt{\gamma}=1+\delta\gamma/2$ and $\gamma^{ij}=\delta^{ij}-\delta\gamma^{ij}$. The combination of extrinsic curvatures on the right-hand side can be re-expressed using the Gauss-Codazzi relation,
\be
\R_{ikjl}|_\Sigma=\R_{ikjl}^{\Sigma} - K_{kj}^\ha K_{\ha\,il} + K_{kl}^\ha K_{\ha\,ij}~,
\label{gc}
\ee
where $\R_{ikjl}^{\Sigma}$ is the intrinsic curvature of the entangling surface. In particular, we obtain
\bea
 \delta S&=& {c\over 6\pi} \int d^2y \sqrt{\gamma} \big(\gamma^{ij}\delta^{ac}\R_{i\ha \hc j} +\delta^{\ha\hc} \, \delta^{bd}\R_{a b c d}
 +\gamma^{ij}\gamma^{kl}\R_{ikjl} - \R^\Sigma \big) \log(\ell/\delta) 
 \non
 &=& {c\over 2\pi} \int d^2y \sqrt{\gamma} \, \delta^{ac} \, \delta^{bd} C_{a b c d}\, \log(\ell/\delta) ~,
 \label{Suniv}
\eea
where the topological term $\int\mathcal{R}^\Sigma$ was discarded in the last equality since it is insensitive to perturbations, and we also made use of the definition of the Weyl tensor, 
\be
C_{a b c d} = \frac{1}{3}\(\delta^{ac}\delta^{bd}\R_{abcd}
 + \gamma^{ij}\delta^{a c} \R_{i a c j} +  \gamma^{ij}\gamma^{kl} \R_{ikjl}\)~.
\ee

Our result \reef{Suniv} should be compared with Solodukhin's general expression for universal entanglement entropy across any surface for a four-dimensional CFT \cite{solo}
\be
 S={1\over 2\pi} \int d^2y \sqrt{\gamma}\[ c \, ( \delta^{\ha\hc}\delta^{\hb\hd}C_{\ha\hb\hc\hd}+K^\ha_{ij} K^{ ij}_\ha-{1\over 2}K^\ha K_{\ha})-a \, \R^\Sigma \] \log(\ell/\delta) ~,
 \labell{eq:solo}
\ee
where the last term in (\ref{eq:solo}) is topological and, since it is insensitive to perturbations, can be ignored in comparing with (\ref{Suniv}). 

Clearly, there is discrepancy at second order between (\ref{Suniv}) and (\ref{eq:solo}), with (\ref{eq:solo}) having additional extrinsic curvature terms.  A few comments:
\begin{itemize}

\item  Eq.~(\ref{Suniv}) was derived on general grounds, with essentially the only assumption being that universal entanglement entropy can be written as a perturbative expansion. Eq.~(\ref{Suniv}) was found through consideration of all possible contractions of the metric perturbation $h_{\mu \nu}$ consistent with symmetries. Demanding the result be reparameterization invariant along the entangling surface imposed constraints (\ref{eq:cons}) on the coefficients of the possible contractions. 
After integration over the transverse space, the relative coefficients of the terms $\R_{iacj}\delta^{ij} \delta^{ac}$, $K_a K^a$, and $K_{ij}^a K_a^{ij}$ are completely fixed. 
The result, however, is not in full agreement with (\ref{eq:solo}).
\item Two of the coefficients, $\alpha_1$ and $\alpha_2$, are not fixed by any consistency conditions. Nevertheless, an explicit first order calculation \cite{RS1} reveals that they are in agreement with (\ref{eq:solo}). It is therefore interesting that (\ref{Suniv}) and (\ref{eq:solo}) agree at first, but not at second, order. It is even more intriguing that (\ref{Suniv}) is Weyl invariant.
\item While the arguments for (\ref{Suniv}) appear robust, there is at the same time a great deal of evidence for (\ref{eq:solo}). Eq.~(\ref{eq:solo}) was originally found \cite{solo} by demanding that the universal part of entanglement entropy be Weyl invariant, \footnote{This assumption follows from the fact that in the replica-trick, universal entanglement entropy is computed from the expectation value of the trace of the energy-momentum tensor on a conifold or equivalently from the anomalous (local) part of the effective action \cite{solo,Ryu:2006ef,Myers:2010tj}. As this is given by a Weyl invariant combination of curvatures, the same holds for the universal entanglement entropy.} and combining that with use of  Ryu-Takayanagi  to fix some of the coefficients in (\ref{eq:solo}). Eq.~\reef{eq:solo} was later rederived through squashed cone techniques \cite{Fur13}. Furthermore, (\ref{eq:solo}) has undergone holographic  \cite{Hung:2011xb} \footnote{In particular, \cite{Hung:2011xb} tested (\ref{eq:solo}) through use of Gauss-Bonnet gravity in the bulk, with the Jacobson-Myers entropy functional acting as the corresponding holographic entanglement entropy.} and numerical \cite{Huerta:2011qi} tests.

\end{itemize}

\section{Discussion} \label{sec:discussion}
This paper has continued the approach of perturbatively computing entanglement entropy within quantum field theory. The starting point of the perturbative expansion is theories and entangling surfaces for which the reduced density matrix is known. For instance, one begins with a CFT and a planar entangling surface in flat space. One then computes the entanglement entropy for a QFT which is a relevant deformation of the CFT, and for an entangling surface that is slightly deformed and in a weakly curved background. The computation relies on perturbatively expanding the action appearing within the path integral defining the reduced density matrix, and correspondingly perturbatively expanding the von Neumann entropy of the reduced density matrix. In Sec.~\ref{sec:rel}, we found the universal entanglement entropy arising from a relevant perturbation to a CFT, up to second order in the coupling (see Eq.~\ref{main}).  In Sec.~\ref{sec:geom},  we studied the universal entanglement entropy arising from geometric deformations. 
Several puzzles remain.

\subsubsection*{Geometric perturbations at second order}
Most pressing is the tension between the form of the result a perturbative calculation for geometric perturbations must give (\ref{Suniv}) and Solodukhin's expression (\ref{eq:solo}). The extensive checks that (\ref{eq:solo}) has undergone leads us to believe that the perturbative calculation is missing something. Yet, at the same time, while an explicit second order calculation for geometric perturbations involves many subtleties and has room for error, the arguments leading to (\ref{Suniv}) are far more general. Indeed, essentially the only thing assumed to find (\ref{Suniv}) is that entanglement entropy can actually be computed perturbatively via (\ref{Staylor}). For instance, in Sec.~\ref{sec:geom} we did not even assume any particular form for the relation between $\delta S$ and $\delta \rho$. While one can certainly choose to question the validity of any perturbative calculation of entanglement entropy for a deformed plane, it would be odd that the first order result matches (\ref{eq:solo}). It may be that for the second order calculation, we should relax the assumptions made in (\ref{sec:geom}) on the general form of the contact term (as described in Footnote \ref{footc}). It might also be instructive to perform the second order geometric perturbation computations for Renyi entropies, along the lines of \cite{Lewkowycz:2014jia}.

\subsubsection*{An extra boundary term?} 
It appears to us that the most promising resolution would be that there is an additional boundary term, residing on the entangling surface, that is in some sense ``non-perturbative''. This term would need to be added to the perturbative calculation, so that the perturbative expression (\ref{Suniv}) combined with this new boundary term, yields Solodukhin's expression (\ref{eq:solo}). It is especially intriguing that the perturbative expression (\ref{Suniv}) is just the Weyl tensor, without the additional combination of extrinsic curvatures terms one finds in (\ref{eq:solo}).

More generally, it is an open question if the textbook procedure of computing entanglement entropy as the von Neumman entropy of the reduced density matrix for a subregion is in itself well-defined and unambiguous. In the context of gauge fields, a piece of a Wilson loop cutting across the subregion  is not gauge invariant; the placement of charges on the boundary of the subregion provides a cure  \cite{Donnelly:2011hn, Donnelly:2014gva}.
Indeed, even in the context of other fields, ambiguities arise. To actually compute entanglement entropy, one must introduce a UV-cutoff. Once this is done, one must address how to treat the algebra of observables residing on the boundary \cite{Casini:2013rba, Casini:2014aia} (see also \cite{Ohmori:2014eia}).
In the context of non-minimally coupled scalars, there is again a puzzle. In the continuum,  entanglement entropy is clearly the same for the minimally and non-minimally coupled scalar, as the stress-tensor has no impact on the spectrum of the reduced density matrix. Yet, in the continuum the entanglement entropy is infinite. The  UV cut-off one must impose to regulate it is sensitive to the definition of the stress-tensor. The difference between the modular Hamiltonians for the minimally and non-minimally coupled scalars is a boundary term. Taking this seriously, one finds that the minimally and non-minimally coupled scalar give different  entanglement entropies \cite{Larsen:1995ax, RS3, MRS}. An alternative is to insist that minimally and non-minimally coupled scalars give the same entanglement entropy \cite{Solodukhin:1996jt, Hotta:1996cq, Solodukhin:2011gn, Klebanov:2012va, Nishioka:2014kpa, Lee:2014zaa, Herzog:2014fra, Dowker:2014zwa}. Ref. \cite{Lee:2014zaa} gave some quantitative demonstrations  that the treatment of the region
near the entangling surface is directly correlated with the form of the scalar
modular Hamiltonian. Recently, \cite{Casini} argued that for a free theory, one should use the modular Hamiltonian constructed with the canonical stress-tensor, whereas for an interacting theory one should use the conformal stress-tensor. 

In short, it appears that universal entanglement entropy may be less universal than had been appreciated. While a $\log$ divergence is invariant under a change in the UV-cutoff, the very presence of a UV-cutoff brings the physics of the entangling surface into play. It would be good to understand better and more generally what the correct boundary choices are, and the extent to which they are unique.

\subsubsection*{Contact terms}
The expression for the change in entanglement entropy under a relevant perturbation with operator $\mathcal{O}$ consists of correlation functions with the insertion points of the operators integrated over the space (see Eq.~(\ref{2nd})). A question which needs to be addressed is: should contact terms be included in evaluating this expression? 
In particular, at second order one needs to evaluate $\langle T_{\mu \nu} \mathcal{O} \mathcal{O}\rangle$. A contact term arises if  $T_{\mu \nu}$ collides with $\mathcal{O}$, leading to a correlator of the form $\langle \mathcal{O} \mathcal{O}\rangle$.\footnote{Another possible contact term is if the two $\mathcal{O}$'s collide. Since $\mathcal{O} \mathcal{O} \sim \mathcal{O}$, one then has a term $\langle T_{\mu \nu} \mathcal{O}\rangle$. For a CFT this vanishes. So, at least at second order, this type of contact term does not contribute.}  Thus, the result for the universal part of entanglement entropy is sensitive to the inclusion or exclusion of the contact term. 

Contact terms are in themselves something of an oddity. They arise when a new term needs to be added to a correlation function so that it becomes a well-defined distribution at coincident points (see, for instance, \cite{Osborn:1993cr,Petkou:1999fv, Bzowski:2013sza}). The coefficient of the contact terms is found by demanding correlation functions satisfy certain consistency conditions, such as the Ward identities. Part of the obscurity of contact terms is that they defy a clear physical interpretation.  It is therefore interesting that they may affect entanglement entropy. 

While we would have otherwise expected contact terms should be included \cite{RS3}, in deriving the second order result for entanglement entropy, (\ref{main}), we  in fact \textit{did not} include the contact term. If the contact term were to be included, (\ref{main}) would be replaced by a verifiably incorrect expression for the entanglement entropy. 

Contact terms in the context of geometric perturbations, involving the collision of two energy-momentum tensors, must also  be understood. In this context what is clear is that if one does not include any contact terms, then the resulting expression for entanglement entropy is not even reparameterization invariant.

\acknowledgments  
We thank  D.~Kabat, S.~Leichenauer, D.~Marolf, M.~Mezei, P.~McFadden, R.~Myers, E.~Perlmutter, K.~Skenderis, and M.~Srednicki for helpful discussions. 
The work of VR is supported by NSF Grants PHY11-25915 and PHY13-16748. 
The work of MS is supported in part by NSF Grant PHY-1214644 and the Berkeley Center for Theoretical Physics.

\appendix

 \section{Evaluation of $<K \mathcal{O} \mathcal{O}>$} \label{sec:appendix}
In the appendix we evaluate the integrals appearing in the calculation in Sec.~\ref{sec:relevant} of the $\langle K \mathcal{O}  \mathcal{O}\rangle$ contribution to the change in entanglement entropy under a relevant deformation. We start by deriving an integral identity that will prove useful. First recall that
\be
  \int {d^dk \over (2\pi)^d} {e^{ik\cdot x} \over (k^2)^\al} = {1\over (4\pi)^{d/2}} {\Gamma(d/2-\al)\over \Gamma(\al)} \({x^2\over 4}\)^{\al-d/2}~.
\ee
Using this gives,
\be
 \int d^dx_3 \, {1\over x_{23}^\gamma x_{13}^\bt} = \pi^{d/2} 
 {\Gamma\({d-\gamma \over 2}\)  \Gamma\({d-\bt \over 2}\) \Gamma\({\bt+\gamma-d\over 2}\)\over \Gamma\({\gamma\over 2}\)\Gamma\({\bt\over 2}\)
 \Gamma\(d- {\bt+\gamma\over 2}\)} \, x_{12}^{d-\bt-\gamma}~,
 \labell{ident1}
\ee
where $x_{ij}=x_i-x_j$ for $i,j=1,2,3$. 
Differentiating \reef{ident1} with respect to $x_2$, one can establish an additional  set of identities such as,
\bea
 \int d^dx_3 \, {(x_{32})_\mu(x_{32})_\nu\over x_{23}^\gamma x_{13}^\bt} &=& \pi^{d/2} 
 {\Gamma\({d-\gamma +2 \over 2}\)  \Gamma\({d-\bt \over 2} \) \Gamma\({\bt+\gamma-d-2\over 2}\) 
 \over 4 \Gamma\({\gamma\over 2}\)\Gamma\({\bt\over 2}\) \Gamma\( d- {\bt+\gamma\over 2} + 2\)}
  \, x_{12}^{d-\bt-\gamma+2}
 \non
 &\times& \( (d-\bt) \delta_{\mu\nu}  
  +  (d+2-\gamma)(\bt+\gamma-d-2)   {  (x_{12})_\mu(x_{12})_\nu\over x_{12}^2 }   \) ~.
  \non
  \labell{ident2}
\eea

\subsection*{Free scalar field}
For the computation of the entanglement entropy for the free scalar field we needed to evaluate $\langle K \mathcal{O} \mathcal{O}\rangle$, given by (\ref{eq:KOO1}). Making use of the 3-point function (\ref{eq:TOO}) with $x^{\mu=2}=\bar x^{\mu=2}=0$, (\ref{eq:KOO1}) becomes
\begin{multline}
\langle K \mathcal{O} \mathcal{O}\rangle = - (2\pi)^2 \frac{4}{(d-2) S_d^3}\, \int d^{d-2} y \int_0^\infty dx_1 \, x_1\int d^{d-2} \bar y  \int_{-\infty}^0 d\bar x_1\, \bar x_1  \int d^d z \,
\\
\times \frac{(x_{\rho} - \bar x_{\rho})(x^{\rho} - z^{\rho})}{(x - \bar x)^d\, (x- z)^d\, (\bar x-z)^{d-2}}
\end{multline}
We now rewrite the integrand as follows,
\be
\frac{(x_{\rho} - \bar x_{\rho})(x^{\rho} - z^{\rho})}{(x - \bar x)^d\, (x- z)^d\, (\bar x-z)^{d-2}}=
\, {x_{\rho} - \bar x_{\rho}\over (2-d) (x - \bar x)^{d}(\bar x-z)^{d-2}} \, {\del\over\del x_\rho} \, (x-z)^{2-d}~.
\ee
Using \reef{ident1} to integrate this expression over $z$ gives,
\bea
 \langle K \mathcal{O} \mathcal{O}\rangle &=&  (2\pi)^2 \pi^{d/2} \frac{\Gamma\({d-4\over 2}\)}{\Gamma(d/2)^2 S_d^3}\, \int d^{d-2} y \int_0^\infty dx_1 \, x_1\int d^{d-2} \bar y  \int_{-\infty}^0 d\bar x_1\, \bar x_1 
 \\
 &\times&{x_{\rho} - \bar x_{\rho}\over (x - \bar x)^{d}} \, {\del\over\del x_\rho} \, (x-\bar x)^{4-d}
 \non
 &=& - \frac{2 (2\pi)^2 }{(d-2) S_d^2}\, \int d^{d-2} y \int_0^\infty dx_1 \, x_1\int d^{d-2} \bar y  \int_{-\infty}^0 d\bar x_1\, \bar x_1 
 {1 \over (x - \bar x)^{2(d-2)} }~.
 \nonumber
 \eea
Now we do the integral over $\bar{y}$,
\be
\int d^{d-2} \bar{y} \frac{1}{\((x_1 - \bar{x}_1)^2 + \bar{y}^2\)^{d-2}} = \frac{ S_{d-1}}{2^{d-2}}  \frac{1}{(x_1 - \bar{x}_1)^{d-2}}~.
\ee
Next we note that
\be
\int_{-\infty}^{0} d\bar{x}_1 \, \int_0^{\infty} d x_1 \, \frac{x_1\, \bar{x}_1}{(x_1 -\bar{x}_1)^{d-2}} = \frac{(-1)^{d-3}}{(d-4)(d-3)}\int_{0}^{\infty}{\frac{d x_1}{x_1^{d-5}}}~.
\ee
Introducing UV and IR cut-offs, we have for $\langle K \mathcal{O} \mathcal{O}\rangle$,
\be
\langle K \mathcal{O} \mathcal{O}\rangle= {(-1)^{d-2}\over 2 (4\pi)^{d-3\over 2}} \, {\Gamma\({d-4\over 2}\) \Gamma\({d\over 2}\) \over (d-3) \Gamma\({d-1\over 2}\)} 
\int_{\delta}^{m^{-1}}{\frac{d x_1}{x_1^{d-5}}}~.
\ee
As expected, there is a $\log$ divergence only in $d=6$,
\be 
\langle K \mathcal{O} \mathcal{O}\rangle =  - \frac{\A_{\Sigma}}{18 \pi^2} \log(m\delta) ~.
\ee

\subsection*{CFT}
Here we evaluate $\langle   K \mathcal{  O}\mathcal{  O}\rangle$ for a CFT.  Using \reef{TOO} and \reef{defs}, the integrand in  \reef{KOO} becomes, \footnote{Note that $\bar x_2=x_2=0$.}
\be
  \langle T_{22}(x) \mathcal{O}(\bar x) \mathcal{O}(z) \rangle={a\,z_2^2\over |x-\bar x|^{d-2} |\bar x-z|^{2\Delta-d+2} |z-x|^{d+2}}
  -{a\over d}\,{1\over |x-\bar x|^{d} |\bar x-z|^{2\Delta-d} |z-x|^d}
\ee
Next we rewrite the above expression as,
\bea
  \langle T_{22}(x) \mathcal{O}(\bar x) \mathcal{O}(z) \rangle&=&
  {a\over d(d-2)}{1\over |x-\bar x|^{d-2}} {\del^2\over\del x_2^2} \( {1\over |\bar x-z|^{2\Delta-d+2} |z-x|^{d-2}}\)
  \\
  &+&{a\over d}\,{(x-\bar x)(x-z)-(\bar x-z)(x-z)\over |x-\bar x|^{d} |\bar x-z|^{2\Delta-d+2} |z-x|^d}
  \non
  &=&
  {a\over d(d-2)}{1\over |x-\bar x|^{d-2}} {\del^2\over\del x_2^2} \( {1\over |\bar x-z|^{2\Delta-d+2} |z-x|^{d-2}}\)
  \non
  &-&{a\over d(d-2)}\,{(x-\bar x)^\mu\over |x-\bar x|^{d}} {\del\over\del x^\mu} {1\over |\bar x-z|^{2\Delta-d+2} |z-x|^{d-2}}
  \non
  &-&{a\over d(2\Delta-d)(d-2)}\,{1\over |x-\bar x|^{d}}\,{\del^2\over\del x_\mu\del\bar x^\mu}{1\over  |\bar x-z|^{2\Delta-d} |z-x|^{d-2}}~.
  \nonumber
\eea
Using \reef{ident1}, one can integrate over $z$. The first two terms exactly cancel each other, while the third term results in
\be
\int d^dz \langle T_{22}(x) \mathcal{O}(\bar x) \mathcal{O}(z) \rangle={S_d\over d(d-2\Delta)}\, {a\over (x-\bar x)^{2\Delta}}~
\ee
Substituting this result back into \reef{KOO} and integrating subsequently over $\bar y$, $y$ and $\bar x_1$ leads to,
\be
 \langle  K \mathcal{ O}\mathcal{ O}\rangle={4\,a\,\pi^{d+1}\over d (d-2\Delta)(d-2\Delta-1)}\,{\Gamma(\Delta-d/2)\over\Gamma(\Delta)\Gamma(d/2)}
 \A_{\Sigma}\int_\delta^\ell dx_1 x_1^{d-2\Delta+1}~,
\ee
where $\delta$ and $\ell$ are UV and IR cut-offs, respectively.

\section{Relevant perturbations: free fields} \label{appendixB}
In section \ref{sec:relevant} we found the dependence of the universal part of entanglement entropy across a plane for a CFT deformed by a relevant operator, to second order in the deformation (see Eq.~\ref{main}). In this appendix, we check (\ref{main}) for the special case of free fields deformed by a mass term.

\subsubsection*{Fermion}
In $d=4$, the relevant deformation corresponds to a fermionic mass term, $\lambda_\psi=m_\psi$ and $\mathcal{O}_\psi(x)=\bar\psi\psi(x)$. From the massless fermion two-point function 
\be
\langle\psi(x)\bar\psi(0)\rangle={1\over S_d} {\gamma_\mu x^\mu\over x^d}~,
\ee
where $\gamma_\mu$ are the Euclidean gamma matrices, we have
\be
 \langle \mathcal{O}_\psi(x) \mathcal{O}_\psi(0)\rangle = {2^{\[{d\over2}\]}\over S_d^2} {1\over x^{2(d-1)}}\,.
\ee
Comparing with \reef{2pO}, we deduce $ N_\psi={2^{\[{d\over2}\]}\over S_d^2}$.
Substituting this normalization constant into \reef{main}, we get for Dirac fermions in $d=4$,
\be
 \delta S= {m^2_\psi\A_{\Sigma} \, \over 12\pi }\log(m_\psi\delta)~,
\ee
in agreement with the literature \cite{Huerta:2011qi,Lewkowycz:2012qr}. 

\subsubsection*{Conformal Scalar}
Similarly, one can consider a conformally coupled scalar in $d=6$ that is deformed by a mass term. In this case, $\lambda_\phi=m_\phi^2/ 2$ and $\mathcal{O}_\phi(x)=\phi^2(x)$. Comparing the correlator 
\be
\langle \mathcal{O}_\phi(x) \mathcal{O}_\phi(0)\rangle={2\over (d-2)^2 S_d^2} {1\over x^{2(d-2)} }
\ee
with \reef{2pO}, we deduce $ N_\phi={2\over (d-2)^2 S_d^2}$ and thus find, 
\be \label{eq:Sconf4}
 \delta S=   {m^4_\phi\A_{\Sigma}\over 960 \, \pi^2}\log(m_\phi\delta)~.
\ee
This is in agreement with the result for the conformally coupled scalar found in \cite{MRS}. There, using the methods of \cite{RS3}, one solves the entanglement flow equation (\ref{relflow}) by computing the correlator $\langle K \mathcal{O}\rangle$ for the \textit{massive} theory, thereby giving a nonpertrubative expression for the entanglement entropy. 

Note that, as follows from (\ref{relflow}), the minimally and conformally coupled scalar have different entanglement entropies as they have different stress-tensors \cite{RS3, MRS} (for a different interpretation, see \cite{Lee:2014zaa}). In particular, the stress-tensor of a conformally coupled scalar is given by 
\begin{equation} \label{eq:Tscalar2}
T_{\mu \nu}^{0} = \partial_{\mu} \phi \partial_{\nu} \phi - \frac{1}{2} \delta_{\mu \nu} (\partial \phi)^2 + \xi_c(\delta_{\mu \nu} \partial^2 - \partial_{\mu} \partial_{\nu})\phi^2~,
\end{equation}
where $\xi_c =(d-2)/4(d-1)$ is the conformal coupling. In Sec.~\ref{sec:scalar} we computed the entanglement entropy for the minimally coupled scalar, reproducing the standard result in the literature. If one repeats the calculation of Sec.~\ref{sec:scalar} using the stress-tensor (\ref{eq:Tscalar2}), one finds (\ref{eq:Sconf4}).

\section{Geometric perturbations: second order terms} \label{AppendixC}
In Sec.~\ref{sec:geom}, the form of the second order contribution to the universal entanglement entropy across a deformed planar entangling surface was found on general grounds. In this appendix, we explicitly compute the second order terms.

Our notation and setup follow \cite{RS1}, where the first order terms were found. Under a small change in the reduced density matrix, $\rho \rightarrow \rho + \delta \rho$, the entanglement entropy, $S = -\text{Tr} \rho \log \rho$, undergoes a change, \footnote{Eq.~\reef{eq:FirstLaw} was derived in \cite{Marolf:2003sq, Bhattacharya:2012mi,Blanco:2013joa, Wong:2013gua} and bears the name of the first law of entanglement entropy. In \cite{Bhattacharya:2012mi,Blanco:2013joa, Wong:2013gua} the Hilbert space is fixed and the variation of the density matrix results from,\eg excited states close to the vacuum, $|0\rangle +\epsilon |\psi\rangle$. In our case the Hilbert space is in general not fixed. In particular, the vacuum state before and after the variation is not the same. This in turn results in a nontrivial $\delta\rho$. Hence, \reef{eq:FirstLaw} is just a generalized form of the first law of entanglement entropy, which also accounts for variations in the Hilbert space.}
\be \label{eq:FirstLaw}
\delta S= \text{Tr}(\delta \rho K) + O(\delta \rho^2)~.
\ee 
For geometric perturbations up to second order, we will only need the term shown in (\ref{eq:FirstLaw}).~\footnote{We ignore $\mathcal{O}(\delta\rho^2)\sim\langle T T \rangle$ terms since they would not give rise to logarithmically divergent terms in this case; see the end of Appendix \ref{sec:general} for details.} Next, we need to find $\delta \rho$ resulting from the perturbation $h_{\mu \nu}$. Expanding the action gives,
\be \label{eq:Iexp}
I = I_0 -\frac{1}{2} \int  T^{\mu \nu} h_{\mu \nu} + \frac{1}{2}\int \int \frac{\delta^2 I}{\delta  g_{\mu \nu} \delta g_{\alpha \beta}}h_{\mu \nu} h_{\alpha \beta} + \ldots~,
\ee
where we made use of the definition of the stress-tensor. Inserting (\ref{eq:Iexp}) into the path integral definition of the reduced density matrix  yields, 
\begin{multline} \label{eq:deltarhoG}
\delta \rho = {1\over \mathcal{N}} 
 \int_{\phi(\C_+)=\phi_+  \above 0pt \phi(\C_-)=\phi_- } \mathcal{D}\phi \,e^{-I_0} 
  \(\frac{1}{2}\int T^{\mu \nu} h_{\mu \nu} + \frac{1}{8} \int \int T^{\mu \nu} T^{\alpha \beta} h_{\mu \nu} h_{\alpha \beta} \right.
  \\
  \left.-\frac{1}{2} \int \int \frac{\delta^2 I}{\delta  g_{\mu \nu} \delta g_{\alpha \beta}}h_{\mu \nu} h_{\alpha \beta} \) + \ldots~.
\end{multline}
where $\phi$ collectively denotes all the fields, $\mathcal{N}$ is a normalization constant, $\C_\pm$ are the two sides  of  $(d-1)$-dimensional cut $\mathcal{C}$ with $\del\,\mathcal{C}=\Sigma$,  and $\phi_{\pm}$ are some fixed field configurations. Substituting $\delta \rho$ into (\ref{eq:FirstLaw}) results in three terms that may contribute at second order,
\bea
 \delta S_1 &\equiv& \frac{1}{2}\int d^dx\, \langle K\, T^{\mu \nu}(x) \rangle h_{\mu \nu}(x)~,
 \non
  \delta S_2 &\equiv& \frac{1}{8} \int d^dx \int d^dx'  \langle K\ T^{\mu \nu}(x) T^{\alpha \beta}(x') \rangle\, h_{\mu \nu}(x) h_{\alpha \beta}(x')~,
 \label{terms} \\
   \delta S_3 &\equiv& -\frac{1}{2} \int d^dx \int d^dx' \langle K \frac{\delta^2 I}{\delta  g_{\mu \nu}(x) \delta g_{\alpha \beta}(x')} \rangle h_{\mu \nu}(x) h_{\alpha \beta}(x')~,
  \nonumber
\eea
and we analyze each in turn.

The $\delta S_1$ term is what was used in the first order calculation \cite{RS1}. Here we use it to account for the second order terms in $h_{\mu \nu}$. As follows from \reef{h}, there is only one such term $h_{i j} \supset x^{a}x^{c}\, K_{c\ i l} K_{a j}^{l}$. Now using \cite{RS1,conifold},
\be \label{eq:TK}
\langle  K T_{i j}(x)\rangle = \frac{c}{3 \pi^2} \frac{\delta_{i j}}{r^4}~, \quad \langle  K T_{i a}(x)\rangle =0~, \quad 
\langle  K T_{ac}(x)\rangle = \frac{c}{3 \pi^2} \frac{4 x_a\, x_c - 3 \delta_{ac}\, r^2}{r^6}~,
\ee
where $r$ is the distance between $x$ and $\Sigma$, we find that \cite{Lewkowycz:2014jia},
\begin{equation} \label{eq:firstsecond}
\delta S_1 = \frac{c}{6 \pi}  \int{d^2y\, \(K_{i j}^a K_{a}^{i j}\)\,  \log(\ell/\delta)}~,
\end{equation}
and 
\be
 \al_1={c\over 6\pi^2}~, \quad \al_2=-3\al_1~, \quad \al_3=4\al_1~.
\ee

Consider now $\delta S_2$ in \reef{terms}. Unless $x$ and $x'$ coincide, $h_{\mu \nu} (x)$ and $h_{\alpha \beta}(x')$ will be at different points, and the expression for $\delta S_2$ will not be local. For instance, if $x\neq x'$, then the curvature tensor $\R_{iajc}$ and the correction to the induced metric $\delta\gamma_{ij}$ in \reef{hh} are evaluated at different points on the entangling surface. As a result, integrating out the transverse space leaves us with a double integral over the insertion points of  $\R_{iajc}$ and  $\delta\gamma_{ij}$. Such a term is not local by definition. Hence, the only way to generate a local contribution to the entanglement entropy is to consider the contact term in the three-point function $\langle K  T_{\mu \nu}(x) T_{\alpha \beta}(x') \rangle$ which identifies $x$ with $x'$ . 

The contact term associated with the merger of two stress-tensors was found by Osborn and Petkou \cite{Osborn:1993cr},
\be
 \mathbb{T}_{\mu\nu}(x) \mathbb{T}_{\al\bt} (x') = \(\frac{C}{c}\, {\pi^4 S_d\over 40}\  h_{\mu \nu \alpha \beta \sigma \rho }^5 + \frac{1}{d}\ (\delta_{\mu \nu}\ h_{\alpha \beta \sigma \rho }^3 + \delta_{\al \bt}\  h_{\mu \nu  \sigma \rho}^3) \) \delta(x-x') \mathbb{T}_{\sigma\rho}(x) + \ldots ~,
 \label{contact}
\ee
where $C$ is a linear combination of the three parameters that define the 3-point correlator of the energy-momentum tensor in a general CFT, $c$ is defined in \reef{trace}, and the ellipses encode terms which will not be relevant for us.
\footnote{The contact term contribution from (\ref{contact}) is in accord with one of the conclusions made in \cite{Lewkowycz:2014jia}. Their analysis indicates that only $\Delta=4$ operators which appear in the OPE of two stress-tensors and have a non-vanishing one-point function in the conical spacetime seem to contribute to the logarithmic divergence of the second order perturbation. To linear order in the deficit angle, such a one point function is a correlator $\langle \mathcal{O} K\rangle$ with $K\sim T_{22}$ in a spacetime without a conical singularity \cite{conifold} . In particular, if the theory is conformal then only $\mathcal{O}\sim T_{\mu\nu}$ generates a non-trivial contribution to the universal entanglement entropy. 
}
In addition,  
\begin{eqnarray}
h_{\mu \nu \sigma \rho}^3 &=& \delta_{\mu \sigma} \delta_{\nu \rho} + \delta_{\mu \rho} \delta_{\nu \sigma} - \frac{2}{d} \delta_{\mu \nu} \delta_{\sigma \rho} ~, \non
h_{\mu \nu \alpha \beta \sigma \rho }^5 &=& \delta_{\mu \sigma} \delta_{\nu \alpha} \delta_{\rho \beta} + (\mu\leftrightarrow \nu, \sigma \leftrightarrow \rho, \alpha \leftrightarrow \beta) \\
&-& \frac{4}{d} \delta_{\mu \nu} h_{\sigma \rho \alpha \beta}^3 - \frac{4}{d} \delta_{\sigma \rho} h_{\mu \nu \alpha \beta}^3 - \frac{4}{d} \delta_{\alpha \beta} h_{\mu \nu \sigma \rho}^3 - \frac{8}{d^2}\delta_{\mu \nu} \delta_{\sigma \rho} \delta_{\alpha \beta}~.
\nonumber
\end{eqnarray}
We have used $\mathbb{T}_{\mu\nu}$ to denote the energy-momentum tensor appearing in \cite{Osborn:1993cr}, as their definition is slightly different from ours, 
\be
  \langle \mathbb{T}_{\mu_1\nu_1}(x_1) \cdots \mathbb{T}_{\mu_n\nu_n}(x_n)\rangle\equiv 
 (-2)^n {\delta\over \delta g^{\mu_1\nu_1} (x_1)}\cdots{\delta\over \delta g^{\mu_n\nu_n} (x_n)} W
  ~,
\ee
where $W$ is the effective action. Using this definition, we can combine $\delta S_2$ with $\delta S_3$,
\bea
 \delta S_2+\delta S_3&=& \frac{1}{2 \, \mathcal{N}} \int d^dx \int d^dx' \, h_{\mu \nu}(x) h_{\alpha \beta}(x') \int \mathcal{D}\phi  \, K \,
 \frac{\delta^2 }{\delta  g_{\mu \nu}(x) \delta g_{\alpha \beta}(x')} \exp({-I(\phi)}) 
 \non
 &=& \frac{1}{8} \int d^dx \int d^dx' \, h_{\mu \nu}(x) h_{\alpha \beta}(x') \langle K \,\mathbb{T}^{\mu\nu}(x)\mathbb{T}^{\al\bt}(x')\rangle
 ~,
\eea
where in the last line we suppressed terms which emerge in the situation when the support of $K\sim T_{\mu\nu}$ overlaps with $\mathbb{T}_{\mu\nu}(x), \mathbb{T}_{\al\bt}(x')$ since such terms do not contribute to the universal entanglement entropy.\footnote{A constant term which emerges if $K\sim T_{22}$ and $\mathbb{T}_{\mu\nu}(x), \mathbb{T}_{\al\bt}(x')$ overlap does not contribute to the universal entanglement entropy, whereas $\langle T T\rangle$ which emerges if only one of the energy-momentum tensors collides with $K$ does not have a non-trivial delta function which may generate a logarithmically divergent term.} 

Substituting \reef{contact}, yields
\bea
 \delta S_2+\delta S_3&=& \(\frac{C}{c}\, {\pi^4 S_d\over 320}\  h_{\mu \nu \alpha \beta \sigma \rho }^5 + \frac{1}{8 d}\ (\delta_{\mu \nu}\ h_{\alpha \beta \sigma \rho }^3 + \delta_{\al \bt}\  h_{\mu \nu  \sigma \rho}^3) \) 
 \non
 &\times& \int d^dx  \langle K \,\mathbb{T}^{\sigma\rho}(x) \rangle  \, h^{\mu \nu}(x) h^{\alpha \beta}(x) ~,
\eea
Using now the tracelessness of $\langle K \,\mathbb{T}^{\sigma\rho} \rangle$ and symmetries of  $h_{\mu \nu \alpha \beta \sigma \rho }^5$, $h_{\alpha \beta \sigma \rho }^3$ and $h_{\mu\nu}$, the above expression can  be simplified,
\bea
 \delta S_2+\delta S_3&=& \(\frac{C}{c}\, {\pi^4 S_d\over 40}\, 
 h_{\sigma\alpha}(x) h^\alpha_\rho(x)+ \frac{1}{2 d}\delta^{\mu \nu}h_{\mu \nu}(x) h_{\sigma\rho}(x) \big(1- \frac{C}{c}\, {\pi^4 S_d\over 10} \big) \) 
 \non
 &\times& \int d^dx  \langle K \,\mathbb{T}^{\sigma\rho}(x) \rangle   ~.
 \label{eq:term2}
\eea

The result for the universal entanglement entropy at second order is thus given by the sum of (\ref{eq:firstsecond}) and (\ref{eq:term2}). However, substituting $d=4$, using \reef{eq:TK} and noticing that in flat space $\langle K \,\mathbb{T}^{\sigma\rho}(x) \rangle=-\langle K \,T^{\sigma\rho}(x) \rangle$, it can be seen that the final expression will suffer from a number of pathologies. First, it will depend on the gauge field $A_i$. Second, the result is not even of the form discussed in Sec.~\ref{sec:geom} as being necessary for an expression that preserves diffeomorphism invariance along the entangling surface.  Indeed, comparing (\ref{eq:term2}) with \reef{Staylor2}, yields
\be
 \al_4=-{c\over 12\pi^2}\big(1- {\pi^6\over 5} \, \frac{C}{c} \big)~, \quad \al_5=-{\pi^4\over 30} \,C ~, \quad \al_6={\al_4\over 2}~, \quad \al_7=2\al_4~,
\ee
which contradicts \reef{eq:cons}. Finally, the contact term \reef{contact} introduces a new parameter $C\neq c$. \footnote{This may indicate that  the contact terms we are using are not the appropriate ones (see also appendix \ref{sec:general}). In particular, the contact terms in the path integral picture should be related to commutators in the Heisenberg picture. This method of establishing the contact terms might not be the same as the one normally used, which is through the Ward identities.} All of these things in themselves indicate the need for additional contributions.

\section{Self-consistency conditions}
\label{sec:general}

The modular Hamiltonian is known for a planar entangling surface for any QFT, and a spherical entangling surface for a CFT \cite{Casini:2011kv}. For other entangling surfaces, little is known about the modular Hamiltonian except that it is nonlocal \cite{Lewkowycz:2014jia}. In this appendix, we  derive a set of self-consistency conditions that follow from the assumption that the density matrix is normalized, $\text{Tr}_{\mt V} \, e^{-K}=1$. These relations give intriguing hints about the structure of the modular Hamiltonian for general entangling surfaces. 

Promoting the coupling constant to an external field $\lambda(x)$ and differentiating the normalization constraint yields,
\be
 0= {\delta\over \delta \lambda(x)}\text{Tr}_{\mt V} e^{-K} = - \text{Tr}_{\mt V} \big( e^{-K} {\delta K \over \delta \lambda(x)} \big) = - \langle {\delta K \over \delta \lambda(x)} \rangle ~,
 \label{norm1}
\ee
where the second equality follows from cyclicity of the trace, whereas the vacuum expectation value on the right side follows from the assumption that the field theory resides in the vacuum.  Differentiating once more yields,
\be
0={\delta\over\delta\lambda(y)}\langle {\delta   K \over \delta\lambda(x)}\rangle \quad \Rightarrow \quad \langle {\delta^2   K \over \delta\lambda(x)\delta\lambda(y)}\rangle=\langle \mathcal{O}(y) {\delta   K \over \delta\lambda(x)}\rangle ~.
\ee
Similarly,
\be
 0=  \langle {\delta K \over \delta g_{\mu\nu}(x)} \rangle  \quad \Rightarrow \quad 
 \langle {\delta^2   K \over \delta g_{\mu\nu}(x)\delta g_{\al\bt}(y)}\rangle = - {1\over 2}\langle \widetilde T^{\al\bt}(y) \, {\delta   K \over \delta g_{\mu\nu}(x)}\rangle ~,
 \label{norm2}
\ee
where for brevity we used the  definition,
\be
\widetilde T^{\mu\nu}(x)\equiv\sqrt{ g(x)}\,T^{\mu\nu}(x) = - 2 {\delta I \over\delta g_{\mu\nu}(x)} \,~.
\ee
Combined with \reef{norm1}, this identity can be used to derive the following relation:
\bea
 \langle \mathcal{O}(y)  {\delta  K\over \delta g_{\mu\nu}(x)} \rangle &=&  - {\delta \over \delta\lambda(y)} \langle {\delta  K\over \delta g_{\mu\nu}(x)} \rangle 
 +\langle {\delta^2  K\over \delta g_{\mu\nu}(x) \delta\lambda(y)} \rangle
 \non
 &=& {\delta\over \delta g_{\mu\nu}(x)}\langle {\delta  K\over \delta\lambda(y)} \rangle
 -{1\over 2}\langle {\delta  K\over \delta\lambda(y)} \widetilde T^{\mu\nu}(x) \rangle
 =-{1\over 2}\langle {\delta  K\over \delta\lambda(y)} \widetilde T^{\mu\nu}(x)  \rangle~,
 \label{self2}
\eea
where \reef{norm1} and  \reef{norm2} were used in the third and second equalities, respectively. 

If the unperturbed entangling surface exhibits rotational symmetry in the transverse space, then the modular Hamiltonian is linear in the energy-momentum tensor, and ${\del  K\over \del\lambda}=\mathcal{O}$ holds by a direct computation. In this case \reef{self2} takes the form
\be
 \langle \mathcal{O} \, {\delta  K\over \delta g_{\mu\nu}(x)} \rangle = -{1\over 2}\langle \mathcal{O} \, \widetilde T^{\mu\nu}(x) \rangle~.
\labell{magic}
\ee
Note that the variation of the modular Hamiltonian on the left hand side of \reef{magic} is completely general, and thus cannot be explicitly carried out without knowing the specific form of $K$. Yet, the right-hand side is just the standard correlator in a QFT, which vanishes if the field theory is conformal. 

Moreover, let us consider an alternative representation of \reef{relflow}
\be
 {\delta S\over \delta\lambda(x)}={ \delta\over \delta \lambda(x)} \text{Tr}_{\mt V} \big( e^{-K} K \big) =  \text{Tr}_{\mt V} \big( e^{-K} { \delta K\over  \delta \lambda(x)} \big) - \text{Tr}_{\mt V} \big( e^{-K} { \delta K\over  \delta \lambda(x)} K \big)
 = -\langle { \delta K\over \delta \lambda(x)} K \rangle ~,
 \label{relflow2}
\ee
where \reef{norm1} and cyclicity of the trace have been used. Comparing with \reef{relflow} yields,
\be \label{magic22}
 \langle { \delta K\over \delta \lambda(x)} K \rangle = \langle \mathcal{O}(x) K \rangle~,
\ee
and similarly,
\be
 \langle { \delta K\over \delta g_{\mu\nu}(x)} K \rangle = - {1\over 2}\langle \widetilde T^{\mu\nu}(x) K \rangle~.
 \label{magic2}
\ee

Equations (\ref{magic22}) and (\ref{magic2}) are as far as we will be able to get. Our findings \textit{suggest} that $\delta  K / \delta \lambda$ and $\delta  K /\delta g_{\mu\nu}$ can be replaced with $\mathcal{O}$ and $-\widetilde T^{\mu\nu}/2$ within any connected correlator in general, and in \reef{2nd} in particular.\footnote{Note, however, that it would clearly be incorrect to identify ${\del K\over \del\lambda}=\mathcal{O}$, as such an identification holds up to a non-local functional of the metric and coupling constants (which drops out of any connected correlator).} However, we do not have a proof to believe such a replacement is generally justified, although it is worth noting that the above identities hold for any  state, provided one uses the appropriate modular Hamiltonian for that state. 
If one \textit{assumes} such a replacement is true, then it can be used to find a closed form expression for the second order term in a perturbative expansion of entanglement entropy for a deformed geometry. 

Indeed, differentiating \reef{eq:Sdef} with respect to the metric results in,

\bea
 \frac{\delta S }{\delta g_{\mu \nu}(x)} &=& \frac{1}{2} \langle \widetilde T^{\mu\nu} (x)   K   \rangle~,
 \label{eq:deltaSgeom1}
\\
 \frac{\delta^2 S }{\delta g_{\al\bt}(y)\delta g_{\mu \nu}(x)} &=& {1\over 4}\langle \widetilde T^{\al \bt} (y) \widetilde T^{\mu \nu} (x)   K   \rangle  
 +\frac{1}{2} \langle {\delta \widetilde T^{\mu \nu} (x)\over \delta g_{\al\bt}(y)}   K   \rangle  
 +\frac{1}{2} \langle \widetilde T^{\mu \nu} (x) {\delta   K  \over \delta g_{\al\bt}(y)}  \rangle  ~.
 \nonumber
\eea
Substituting \reef{eq:deltaSgeom1} into \reef{Staylor} yields,
\begin{multline}
 \delta S=\frac{1}{2} \int d^d x  \, \langle  \tilde T^{\mu\nu} (x)   K   \rangle   h_{\mu\nu}(x)
 +{1\over 8}\int d^d x  \int d^d y \langle  \tilde T^{\al \bt}(y)  \tilde T^{\mu \nu} (x)   K   \rangle   h_{\mu\nu}(x)h_{\al\bt}(y)
 \\
 +{1\over 4}\int d^d x \int d^d y  \( \langle {\delta \widetilde T^{\mu \nu} (x)\over \delta g_{\al\bt}(y)}   K   \rangle   
 + \langle  \widetilde T^{\mu \nu} (x) {\delta   K  \over \delta g_{\al\bt}(y)}  \rangle \)h_{\mu\nu}(x)h_{\al\bt}(y) +\mathcal{O}(h^3_{\mu\nu})~.
 \label{deltaSgeom2}
\end{multline}
with $h$'s defined in \reef{h}. Replacing the derivative of the modular Hamiltonian with $- \widetilde{T}^{\al\bt}(y)/2$ produces a closed form expression for evaluation of entanglement entropy to second order in a given small deformation of the geometry.\footnote{Eq.~ \reef{norm2} provides  additional support for this replacement, since it implies that $\langle T^{\mu \nu} (x) {\delta   K  \over \delta g_{\al\bt}(y)}  \rangle$ is symmetric under $\mu\nu\,, x\leftrightarrow \al\bt \, , y$, and therefore ${\delta   K  \over \delta g_{\al\bt}(y)}\sim T^{\al\bt}(y) \, H$, where $H$ is some scalar operator. However, given \reef{magic} it is unlikely that $H$ is non-trivial.} This is the same expression one obtains through a perturbative expansion of the definition of the von Neumman entropy in terms of the change in the reduced density matrix arising from a change in the action \cite{RS1}.  In particular,  the first 3 terms in this expression are the same as in \reef{terms}, whereas the last term is associated with $\mathcal{O}(\delta\rho^2)$ in \reef{eq:FirstLaw} and does not contribute to the universal part of entanglement entropy. Understanding the assumptions that went into (\ref{deltaSgeom2}) may help in better understanding the treatment of contact terms in the second order computation (Appendix \ref{AppendixC}).


\end{document}